\documentclass[]{tMOP2e}
\usepackage{color}
\usepackage{hyperref}

\begin{document}

\newcommand{\sergei}[1]{{\color{magenta}({\bf Sergei:} #1)}}

\doi{10.1080/0950034YYxxxxxxxx}
 \issn{1362-3044}
\issnp{0950-0340} \jvol{00} \jnum{00} \jyear{2013} \jmonth{10
January}

\markboth{V.R. Tuz, D.V. Novitsky et al.}{Journal of Modern Optics}


\title{Asymmetric bistable reflection and polarization switching \\ in a~magnetic nonlinear
multilayer structure}

\author{Vladimir R. Tuz$^{a,b}$$^{\ast}$\thanks{$^\ast$Corresponding author. Email:
tvr@rian.kharkov.ua \vspace{6pt}}, Denis V. Novitsky$^{c}$, Sergey
L. Prosvirnin$^{a,b}$ and Sergei V. Zhukovsky$^{d}$\\\vspace{6pt}
$^{a}${\em{Institute of Radio Astronomy of National Academy of
Sciences of Ukraine, 4, Chervonopraporna St., Kharkiv 61002,
Ukraine}}; $^{b}${\em{School of Radio Physics, V.N.~Karazin Kharkiv
National University, 4, Svobody Sq., Kharkiv 61022, Ukraine}};
$^{c}${\em{B.I.~Stepanov Institute of Physics of National Academy of
Sciences of Belarus, 68, Nezavisimosti Avenue, Minsk 220072,
Belarus}}; $^{d}${\em{DTU Fotonik -- Department of Photonics Engineering,
Technical University of Denmark,
{\O}rsteds Pl. 343, DK-2800 Kgs.Lyngby, Denmark}}\\\vspace{6pt}\received{v3.4 released October 2010} }

\maketitle

\begin{abstract}
Optical properties of a one-dimensional photonic structure
consisting of Kerr-type nonlinear and magnetic layers under the
action of an external static magnetic field in the Faraday geometry
are investigated. The structure is a periodic arrangement of
alternating nonlinear and magnetic layers (a one-dimensional
photonic crystal) with one of the layers doubled to create a defect
where periodicity is violated. Strong enhancement of nonreciprocity
is observed at the frequencies of the defect modes, where linearly
polarized light incident from one side of the structure undergoes
$90^\circ$ polarization rotation upon reflection, while light
reflected from the other side has its polarization unchanged. Using
the nonlinear transfer matrix calculations in the frequency domain,
it is demonstrated that defect resonances in the nonlinear
reflection spectra undergo bending, resulting in polarization
bistability of reflected light. This bistability is shown to result
in abrupt switching between linear polarization of the output
reflected light when the input intensity is varied. This switching
is confirmed in finite-difference time-domain simulations, and its
hysteresis character is established.
\bigskip

\bigskip
\vspace{12pt}

\end{abstract}

\section{Introduction}

In classical electromagnetism, the reciprocity theorem is based on
the Lorentz lemma \cite{Landau}. It states that the relationship
between an oscillating current and the resulting electric field is
unchanged if one interchanges the points where the current is placed
and where the field is measured; this statement is also called
Helmholtz reversion-reciprocity principle \cite{Born}. In the proof
of the reciprocity theorem it is assumed that the medium where
the field propagates is linear and isotropic. For anisotropic
media, the reciprocity theorem holds only in the case when the
tensors of material parameters of the medium are symmetric. Otherwise,
such as in the case of gyrotropic media (for example, in magnetic
materials), the asymmetry in the tensor of permittivity or permeability makes
 the reciprocity theorem no longer valid.

Nonreciprocity effects are known to be especially pronounced when magnetic
materials are  arranged to form strongly dispersive environments such as photonic crystals,
rather than as bulk materials.
In such environments, nonreciprocity is manifest as spectral asymmetry of dispersion curves
$\omega(\vec k)\ne \omega(- \vec k)$, and particularly arises in
magnetophotonic crystals under the action of an external
static magnetic field in the Faraday geometry.
It was found that in such magnetophotonic crystals \cite{Lyubchanskii,
Inoue} and multilayer systems with structural gyrotropic layers
(featuring light diffraction on a periodic
structure of the medium \cite{Lu:08, Dokukin:09})
a substantial nonreciprocity enhancement is
reached at the edges of the photonic band gaps. However, an even stronger
 localization of the field within the system and, accordingly,
an even more significant nonreciprocity enhancement can be obtained
by purposefully creating one or more defects (resonant cavities) in
a periodic magnetophotonic structure. Using this technique, it is
possible to obtain large values of nonreciprocity with relatively
small values of the external magnetic field,  which facilitates the
use of such structures as optical isolators or one-sided optical
reflectors.


On the other hand, the reciprocity theorem also
does not hold for nonlinear media due to the role of dynamic effects \cite{Landau}.
Therefore, there is an apparent  possibility to obtain new nonreciprocal effects in spatially
inhomogeneous nonlinear media, including ones with dissipative
losses. In such systems,
a number of new nonreciprocal phenomena are revealed, among which spectral nonreciprocity, nonreciprocal
compression \cite{Novitsky_EPL}, and nonreciprocal dynamics of self-induced transparency solitons
\cite{Novitsky_PhysRevA.85.043813} can be mentioned.

As is the case with magnetophotonic crystals, the nonlinearity-induced nonreciprocal response is also
expected to be strongly enhanced in photonic structures comprising an asymmetrically located defect, or cavity, made of a nonlinear (e.g., a Kerr-type) dielectric. This enhancement occurs due to the
strong field localization within the defect layer, which
becomes different for the waves incident on the system from the
opposite sides. Such \textit{spatial-inversion asymmetry} results in a different
coupling strength between the external field and the fields inside the cavity.
This enhancement of nonreciprocal optical response,
sometimes denoted as ``reversible nonreciprocity" \cite{Miroshnichenko}, can be the
basis for designing, for example, optical diodes.

Note that reversible nonreciprocity in an asymmetric nonlinear photonic structure is usually accompanied by optical bistability (or, generally, multistability). This is a general property of any nonlinear system with feedback,
meaning that two or more stable output states may exist for the same input state of the system.
Such bistability can be seen in  magnitude \cite{GrigorievNJP10,Grigoriev2010285}
or polarization \cite{Tuz_JOSAB_2011,Tuz_PhysRevA} of the transmitted and
reflected fields. The relation between input and output
parameters of the system typically forms a hysteresis loop with singularities, where
abrupt switching between the stable states occurs. One can make use of such switching to design active (rather than passive) optical diodes, isolators, and other polarization-control devices.


It is therefore insightful to investigate the effects of the {\em interplay} between
 the Faraday and the Kerr effects in the formation of optical nonreciprocity in
 a photonic structure setting, as well as the influence of such interplay on the dynamics of the bistable switching.
 A straightforward way to do so would be to combine
 a Faraday-active and a Kerr-nonlinear material in the same one-dimensional photonic
 structure with an asymmetrically placed defect. In this paper, we investigate spectral, polarization, and dynamic optical properties of such spatially asymmetric multilayers containing both magnetooptic and nonlinear materials.


By carrying out both frequency-domain (nonlinear transfer matrix)
and time-domain (finite-difference time-domain) calculations, we
show that combining magnetooptic and nonlinear materials in the same
photonic multilayer structure results in many peculiar spectral and
polarization properties. Using a periodic structure with an
asymmetrically placed periodicity violation (a cavity or defect
layer), we show that the nonreciprocal effect can be greatly
enhanced by the interplay between the Faraday effect and the
spatial-inversion asymmetry, leading to a very strong asymmetry in
the polarization conversion via the defect modes even in the linear
regime. Namely, a structure with a good coupling between the field
in the defect and the incident field shows near-$90^\circ$
polarization rotation at the defect resonance frequencies, while
the mirror-image structure (or the same structure when the direction
of incident light is reversed) has poor coupling between the filed
in the defect and the incident field, causing polarization
conversion effects to nearly vanish.

In the nonlinear regime, the bending of resonances in the nonlinear
reflection spectra is shown to lead to strong bistability between
linearly polarized states, again exhibiting strong asymmetry with
respect to the direction of light incidence. This bistability is
shown to give rise to asymmetric polarization switching. Such
switching, along with hysteresis loops in the multilayer
input/output characteristics, is confirmed in time-domain numerical
simulations.


The remainder of this paper is organized as follows. In Section
\ref{sec:theoretical}, we describe in detail the geometrical and
material parameters of the magnetic nonlinear multilayer structure
under study, and outline the nonlinear transfer matrix method that
was used to obtain the spectral properties of the structure. Section
\ref{sec:spectral} follows with the results on these spectral
properties in the linear and nonlinear regime. Strong polarization
conversion, both input/output and polarization bistability
properties are demonstrated. In Section \ref{sec:timedomain}, we
present the numerical simulations results showing bistable switching
between linear polarization states in the time domain. Finally,
Section \ref{sec:conclusions} summarizes the paper.

\section{Problem formulation\label{sec:theoretical}}
\subsection{Magnetophotonic nonlinear multilayers under study \label{sec:structure}}

\begin{figure}[htb]
\centerline{\includegraphics[width=10.0cm]{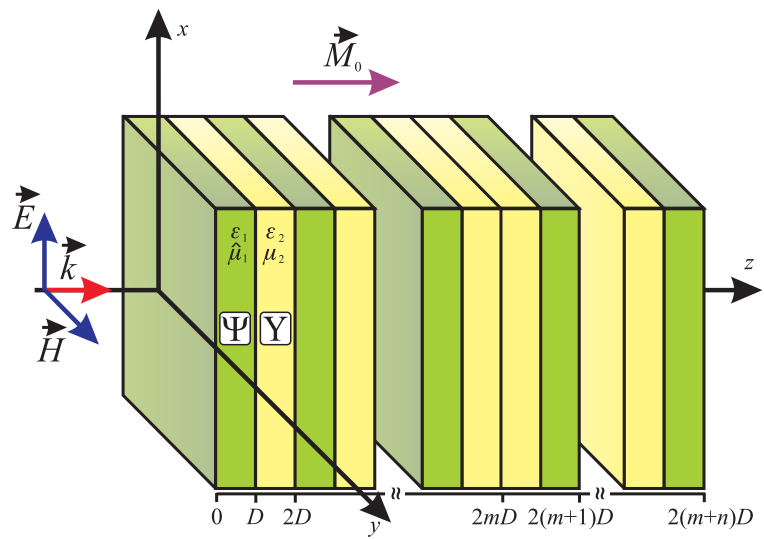}}
\caption{(Color online) Periodic photonic multilayer with a defect containing nonlinear and magnetic
layers in the Faraday geometry.} \label{fig:fig1}
\end{figure}

We consider a planar multilayer stack of infinite transverse extent
(Fig.~\ref{fig:fig1}). Each unit cell is composed of a bilayer which
consists of magnetic layers $\Psi$ (with constitutive parameters
$\varepsilon_1, \hat \mu_1$) and nonmagnetic layers $\Upsilon$ (with
parameters $\varepsilon_2, \mu_2$). The magnetic layers $\Psi$ are
magnetized up to saturation by an external static magnetic field
$\vec M_0$ directed along the $z$-axis (the Faraday geometry). We
assume that each layer $\Upsilon$ is a Kerr-type nonlinear
dielectric, which permittivity $\varepsilon_2$ linearly depends on
the intensity $|E|^2$ of the electric field ($\varepsilon_2=\bar
\varepsilon_2 + \Bar {\Bar\varepsilon}_2 |E|^2$). A periodicity
violation (defect) is created by joining two sections together in
which the layers are alternating in the different order. So, the
defect (repeating two layers of the same type) can be placed either
symmetrically or asymmetrically in the structure. The parameters $m$
and $n$ denote the number of bilayers placed before and after the
periodicity violation. In any case the bilayers are arranged
symmetrically with respect to the defect, i.e. the structure begins
and ends with layers of the same type. We suppose that all layers
have the same thickness $D$. The outer half-spaces $z\le 0$ and
$z\ge \Lambda$ ($\Lambda=2(m+n)D$) are homogeneous, isotropic, and
have material parameters $\varepsilon_0, \mu_0$.

Let us assume that the normally incident field is a linearly polarized
plane monochromatic wave of a frequency $\omega$ and an amplitude
$A$. For the sake of definiteness, we also suppose that the vector
$\vec E$ of the incident wave is directed along the $x$-axis.

A nomenclature of materials which can be used to construct the
proposed structure was previously discussed in \cite{Tuz_PhysRevA}.
It was based on the book \cite{kotov} and reviews on magnetophotonic
crystals \cite{Lyubchanskii, Inoue}. So we suppose that magnetic
layers $\Psi$ can be made from impurity-doped yttrium-iron garnet
(YIG) films, since they are transparent in the near infrared region.
As a material for nonlinear layers $\Upsilon$, either GaAs or InSb
can be selected. Based on \cite{palik}, we expect that nonlinear
response of the proposed structure becomes apparent at the input
light intensity about 1-10~kW/cm$^2$.

In the Faraday geometry when the external static magnetic bias is
parallel to the wave propagation direction ($\vec k\parallel \vec
M_0$), the magnetic permeability $\hat \mu_1$ is a tensor with
off-diagonal components:
\begin{equation}\label{eq:tensor}
\hat \mu_1=
\begin{pmatrix}
\mu_1^T & i\alpha   & 0 \\
-i\alpha  & \mu_1^T & 0 \\
0         & 0       & \mu_1^L
\end{pmatrix}.
\end{equation}

Throughout the paper the working frequencies are supposed to be far
from the frequency of the ferromagnetic resonance of magnetic layers
and band edge if semiconductors such as GaAs are used as nonlinear
materials, so we will assume that losses in the whole structure are
negligibly small. Also, in the chosen frequency band the diagonal
components ($\mu_1^T, \mu_1^L$) of the magnetic permeability
(\ref{eq:tensor}) are close to unity while the off-diagonal ones
($\pm i \alpha$) are  small but non-zero.

\subsection{Nonlinear transfer matrix method}

In the studied structure configuration, in the frequency domain, our
solution is based on the $4 \times 4$-transfer matrix formulation
\cite{vidil} which is used to calculate the inner field distribution
and the reflection and transmission coefficients of the system.
However, this field structure influences the permittivity of the
nonlinear layers through the Kerr effect. Therefore, spectral and
polarization properties of the multilayers under study are
determined using some nonlinear modification of the transfer matrix
method.

Thus, in the \textit{linear regime}, the equation that defines the
coupling of the tangential field components at the input and output
of the magnetophotonic structure is written as
\cite{Tuz_JOSAB_2011, Tuz_PhysRevA}
\begin{equation}\label{eq:matrix}
\vec{V} (0)=\mathfrak{M} \vec{V}(\Lambda)=\left\{( \mathbf{\Psi}
\mathbf{\Upsilon})^m (\mathbf{\Upsilon}
\mathbf{\Psi})^n\right\}\vec{V}(\Lambda),
\end{equation}
where $\vec{V}=\{ E_x,E_y,H_x,H_y\}^T$ is the vector containing the
tangential field components at the structure input and output, the
upper index $T$ denotes the matrix transpose operation, $\Lambda$ is
the total length of the structure; $\mathbf{\Psi}$ and
$\mathbf{\Upsilon}$ are the transfer matrices of the rank four of
the magnetic and nonmagnetic layers, respectively. The elements of
the transfer matrices in Eq.~(\ref{eq:matrix}) are determined from
the solution of the Cauchy problem and are presented in
\cite{vidil}.

Once the solution of the linear problem (\ref{eq:matrix}) is
evaluated, the intensity of the reflected and transmitted fields and
the distribution of the field $\vec {E}_{in}(z)$ inside the system
are calculated. When the structure consists of a Kerr-type nonlinear
layers (in the \textit{nonlinear regime}), the permittivity
$\varepsilon_2$ depends on the intensity of the electric field
within each layer $\Upsilon$ as follows
\begin{equation}
\varepsilon_2(z)=\bar \varepsilon_2 + \Bar {\Bar\varepsilon}_2
|E_{in}(z)|^2. \label{eq:eps}
\end{equation}

Nevertheless, the nonlinearity becomes only apparent within the
defect layers, because here the field is localized most strongly.
Since the thickness of the defect is comparable to the wavelength in
the medium, and hence the intensity variation across the defect
thickness needs to be taken into account, we break up both
$\Upsilon$-layers within the defect into a number of thin sublayers
$\Upsilon_j$ with thicknesses $d_j \ll \lambda$ which are described
with corresponding matrices $\mathbf{\Upsilon}_j$:
\begin{equation}\label{eq:matrix2}
\vec{V} (0)=\mathfrak{M} \vec{V}(\Lambda)=\left\{( \mathbf{\Psi}
\mathbf{\Upsilon})^{m-1}
\mathbf{\Psi}(\mathbf{\Upsilon}_1\ldots\mathbf{\Upsilon}_L)\mathbf{\Psi}
(\mathbf{\Upsilon} \mathbf{\Psi})^{n-1}\right\}\vec{V}(\Lambda).
\end{equation}

We can then consider the dependence of $\varepsilon_2$ on the
average intensity of the electric field $\overline{|E_{in}|^2}$
inside each such sublayer. Knowing the field intensity in each
sublayer $\Upsilon_j$, the actual value of transfer-matrix
$\mathfrak{M}$ of the whole structure can be calculated. Thus we
deal with a system of nonlinear equations on the unknown function of
the field intensity distribution inside the sublayers $\Upsilon_j$.
A magnitude of the incident field $A$ appears as an independent
parameter of this equations set. This system of nonlinear equations
related to the average field intensity distribution in the sublayers
is solved numerically. The solution yields us the final field
distribution in the photonic structure and the values of the
reflection $R$ and transmission $T$ coefficients; the reader is
referred to Ref.~\cite{vidil} for further details.

\section{Spectral properties and polarization bistability \label{sec:spectral}}

In linear optics it is well known that a magnetophotonic structure
in the Faraday configuration has circularly polarized
eigenstates and exhibits dichroism, which leads to the tendency for arbitrary
polarized waves to acquire a polarization state of one of the
eigenwaves of the system as the incoming light propagates through it
\cite{vidil}. In the general case, the result is that the waves become elliptically polarized
after reflection from and transmission through the magnetophotonic structure.
A very important point is
that the spectral dependence of the output polarization state
does not have any discontinuities, and the degree of polarization
transformation depends only on the static magnetic field strength.

However, as we mentioned in the Introduction, the situation is very
different in the nonlinear optical regime. The dependence of
material parameters
on the intensity of light can
lead to polarization bistability, multistability \cite{flytzanis,
Jonsson_PhysRevLett.96.063902, Tuz_JOSAB_2011, Tuz_PhysRevA} or even
to polarization chaos \cite{zheludev}. These effects manifest
themselves in such a way that the output wave polarization state
becomes a multivalued function, containing both stable and unstable branches.

This multivalued character brings about several important
consequences. First, hysteresis loops are known to appear in the
system, both in its input-output characteristic and in its nonlinear
transmission and reflection spectra \cite{ourOptDiodePRA}. Second,
the frequency dependence of the Stokes parameters for the reflected
and transmitted waves shows discontinuous behavior. Third, the
output polarization states may depend on the history of the system,
determined by its temporal evolution. In order to adequately
demonstrate the nonlinear features of the studied structure, both
frequency-domain and time-domain calculations are this warranted.

In this Section, we begin by analyzing the transmission and
reflection spectra of the magnetic nonlinear structure under study.
Since the whole system possesses axial symmetry in the considered
case of normal incidence and Faraday geometry,  we will only
distinguish between copolarized (e.g. $ss$ or $pp$, denoted $co$)
and cross-polarized ($sp$ or $ps$, denoted $cr$) components
throughout the calculations.

\subsection{Linear spectral properties}

\begin{figure}[htb]
\centerline{\includegraphics[width=16.5cm]{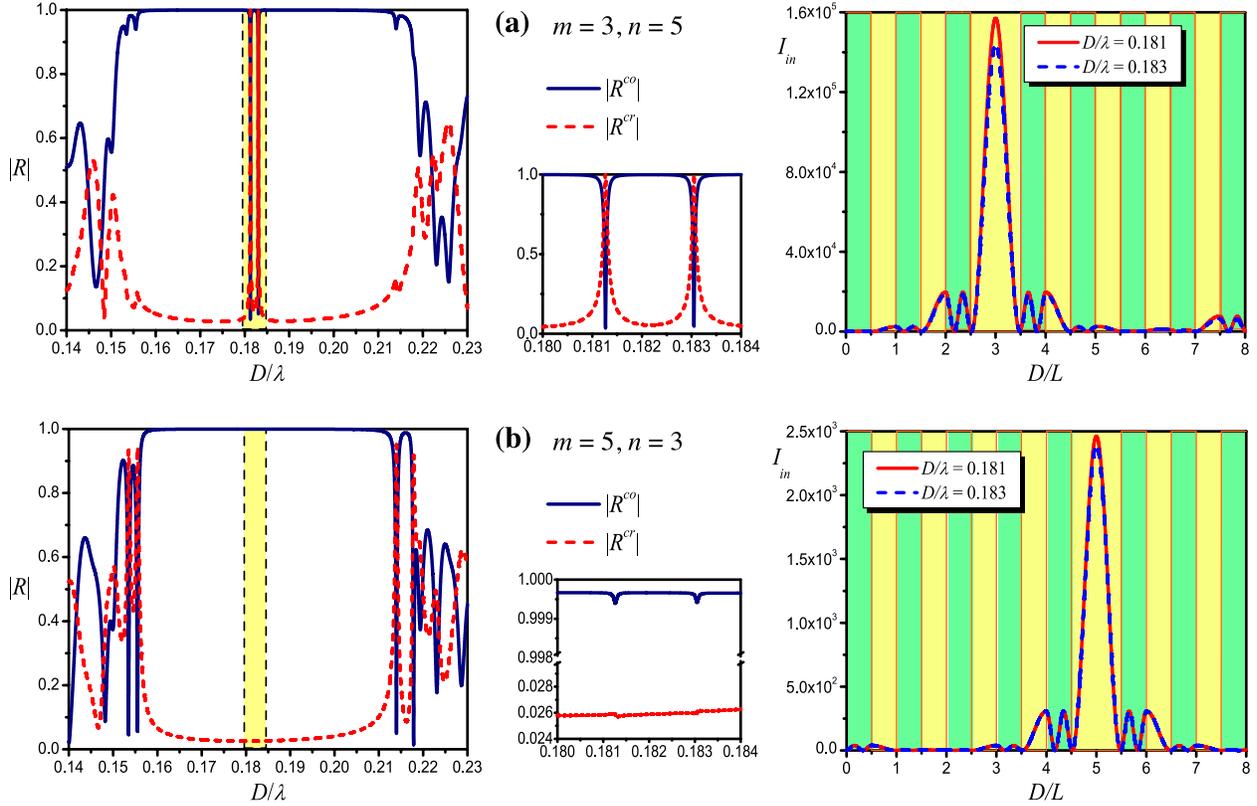}}
\caption{(Color online) Reflection spectra ({\em left}) and field
intensity distributions ({\em right}) for the magnetophotonic
structure with asymmetrically placed defect in the linear regime for
(a) $m=3$, $n=5$ and (b) $m=5$, $n=3$. The spectra show the
magnitudes of the copolarized (co) and cross-polarized (cr)
components of the reflection coefficients for linearly polarized
waves. The insets ({\em middle}) show an enlarged view of the region of interest
($D/\lambda$ from 0.18 to 0.184). Note the difference in the
intensity scale by two orders of magnitude between the field
distribution plots for (a) and (b). The parameters of the structures
are $\varepsilon_1=16$, $\varepsilon_2=2$,
$\mu_1^T=\mu_1^L=\mu_2=1$, $\alpha=0.02$. } \label{fig:fig2}
\end{figure}

In the linear regime the basic optical features of the studied
structure are determined by both the property of its periodicity and
the presence of magnetic layers. As a result of the structure
periodicity, the spectra have interleaved bands with high and low
level of reflection (stopbands and passbands) which are typical for
all photonic structures when the thickness of their layers is
comparable to the wavelength. The corresponding fragment of the
reflection spectra where there is a stopband is presented in
Fig.~\ref{fig:fig2}. If any periodicity violation (a ``defect") is
introduced inside the system, some resonances can appear in the
stopbands, with the field strongly localized inside the defect. The
existence of such localized resonances is explained by the fact that
the periodicity violation forms a resonant Fabry-Perot cavity
enclosed between two Bragg mirrors.

Since the structure consists of magnetic layers under the action of
the external static magnetic field in the Faraday configuration, and
thus its eigenwaves are left-circular-polarized (LCP) and
right-circular-polarized (RCP) waves propagating inside the system
with different speeds \cite{vidil}. Thus the structure reacts
differently to circularly polarized waves with opposite handedness,
which results in distinct resonant conditions for the LCP and RCP
waves. Since the incident linearly polarized wave is a superposition
of two equal-amplitude circularly polarized components of opposite
handedness and different phase, induced by the external static
magnetic field, the localized resonances in stopbands split into
doublets where the position of each peak on the frequency scale is
defined by the corresponding resonant frequency of the RCP and LCP
waves. At the same time, the distribution of the intensity of the
field inside the structure is almost identical on both the resonant
frequencies (Fig.~\ref{fig:fig2}).

It is obvious that if the Bragg mirrors on each side of the cavity
layer have a different number of layers, the properties of the
structure become different for the waves incident on the structure
from the left and right. First of all this difference is manifest in
the intensity magnitude of the field localized inside the defect,
which can be seen in Fig.~\ref{fig:fig2}. From a mathematical point
of view, this property follows from the noncommutativity of the
transfer-matrices product in Eq.~(\ref{eq:matrix}).

This intensity difference strongly  affects on the formation of
localized resonances in the stopbands. When the configuration of the
Bragg mirrors is chosen in such a way that the first mirror has
lower reflectivity ($m=3$) while the second one has higher
reflectivity ($n=5$), the field localization profile inside the
defect is very strong, forming two clearly defined resonances at
$D/\lambda$ between 0.18 and 0.184, where strong polarization
conversion is seen to occur. In the reverse configuration, ($m=5$,
$n=3$), where the highly reflective mirror is between the incident
wave and the defect, the field localization pattern looks quite
similar but the localized field intensity is about 100 times
smaller, and the resonances in the stopband become extremely weakly
pronounced, as seen in the inset of Fig.~\ref{fig:fig2} (b).

It should be pointed out that this strong asymmetry is the result of
an interplay between the geometrical asymmetry of the structure and
the nonreciprocity caused by the Faraday effect. Indeed, in the
absence of magnetic fields the reciprocity theorem would require
that the field localization pattern be similar for both
configurations, and that transmission and reflection peaks
corresponding to a defect resonance be of similar height
\cite{ourPerfectResPRA} for two mirror-symmetric configurations.
However, the Faraday effect violates this condition, which enhances
resonances where there is a good coupling between the incident wave
and the defect ($m=3$, $n=5$).

In one-dimensional photonic crystals, the field intensity
concentrated in the defect layer is known to reach its maximum when
the refractive index of this layer is less than the refractive index
of the remaining structure \cite{wang_PhysStatSol}. Providing this
condition by a proper choice of geometrical and material parameters
of the multilayer, total cross-polarization transformation in the
reflected field can be reached at the defect resonances. Thus, the
near-total linear polarization conversion
\begin{equation}
|R^{co}|=0 \text{ and } |R^{cr}|\approx 1,
\label{eq:totalconversion}
\end{equation}
can be achieved in the ($m=3$, $n=5$) configuration. This can be
seen in the insets of Fig.~\ref{fig:fig2} ({\em middle column})
where the resonance region is highlighted. Conversely, the other
configuration ($m=5$, $n=3$) with poorer coupling between the defect
and the incident wave experiences the reverse effect: the field
localization is weakened and polarization conversion is largely
suppressed.


\begin{figure}[htb]
\centerline{\includegraphics[width=15.0cm]{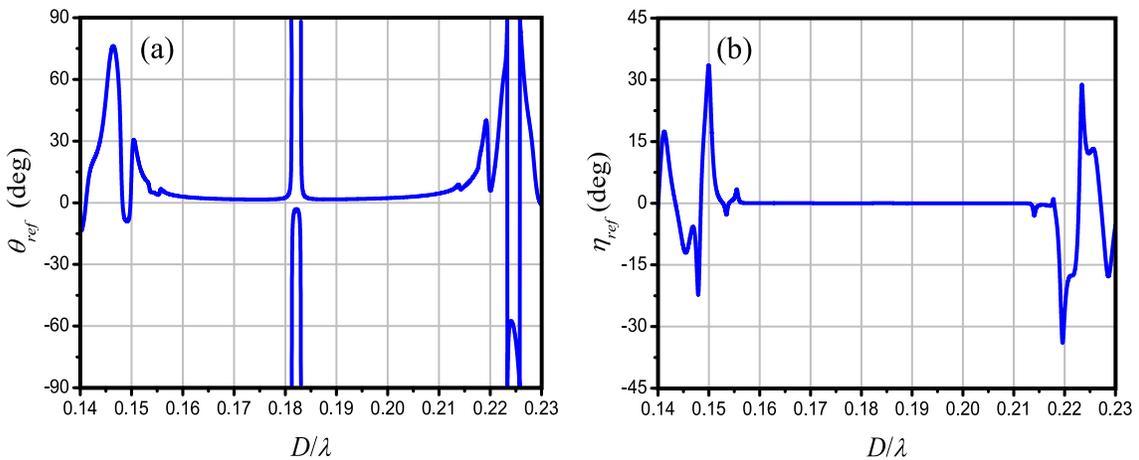}}
\caption{(Color online) Frequency dependences of (a) the
polarization azimuth $\theta$ and (b) the ellipticity angle $\eta$
of the reflected field in the linear regime. The incident light is
linearly polarized, and structure parameters are as in
Fig.~\ref{fig:fig2}.
} \label{fig:fig3}
\end{figure}


Furthermore, the time-reversal asymmetry of the magnetic layers
leads to a change of the polarization state of the reflected field.
This polarization change is confirmed in Fig.~\ref{fig:fig3}, which
shows the corresponding frequency dependences of the polarization
azimuth ($\theta$) and the ellipticity angle ($\eta$) for the
reflected field. According to the definition of the Stokes
parameters, we introduce the ellipticity $\eta$ so that the field is
linearly polarized when $\eta=0$. The case $\eta=-\pi/4$
($-45^\circ$) corresponds to LCP, and the case $+\pi/4$
($+45^\circ$) corresponds to RCP. In all other cases
($0<|\eta|<\pi/4$), the field is elliptically polarized. In the
considered frequency band and in the linear regime, the reflected
field is seen to be linearly polarized throughout the whole stopband
[Fig.~\ref{fig:fig3} (b)], and it acquires a rotation of the
polarization plane on $90^\circ$ exactly at the frequencies of the
localized resonances. This is in full accordance with
Fig.~\ref{fig:fig2} (a).

\begin{figure}[htb]
\centerline{\includegraphics[width=15.0cm]{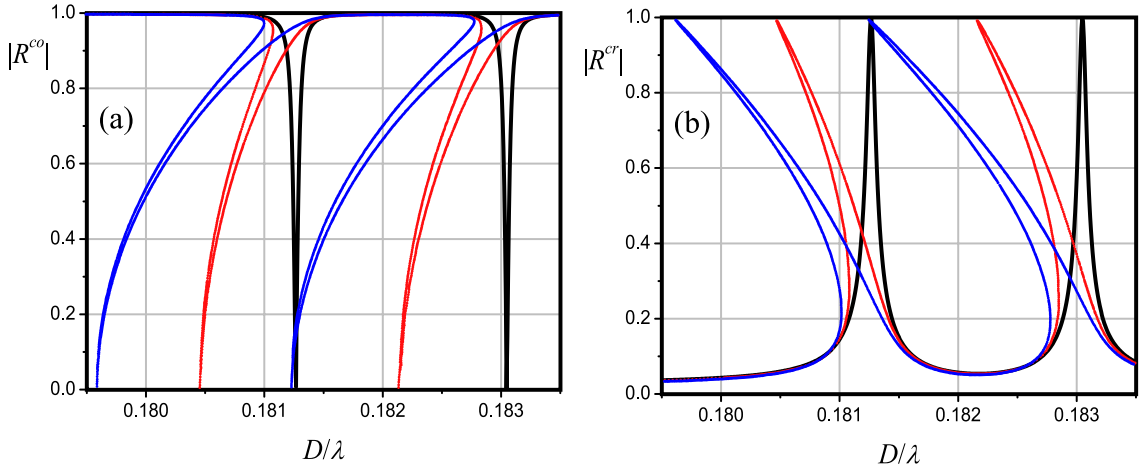}}
\caption{(Color online) Frequency dependences of the magnitudes of
(a) the copolarized (co) and (b) cross-polarized (cr) components of
the reflection coefficients of linearly polarized waves. In the
nonlinear regime the input intensity $I_0$ is taken to be 5 and 10
kW/cm$^2$. Other parameters are as in Fig.~\ref{fig:fig2}.}
\label{fig:fig4}
\end{figure}

\subsection{Nonlinear spectral properties}

In the nonlinear regime the strong field localization in the defect
changes the refractive index of the material within it. In our case
$\Bar {\Bar\varepsilon}_2>0$, so, in the spectral characteristics it
is known to result in bending of both localized resonance in the
stopband towards lower frequencies \cite{ourOptDiodePRA}
(Fig.~\ref{fig:fig4}). Therefore the reflection coefficient
magnitude becomes a multivalued function. The degree of this
resonance bending clearly depends on the intensity of the incident
field and is nearly the same for both resonances in the doublet. Due
to the above-mentioned polarization sensitivity in the
system, a linearly polarized wave will undergo a change in its
polarization state during the reflection also in the nonlinear
regime. However,  the conditions of Eq.~(\ref{eq:totalconversion})
can still be met.

\begin{figure}[htb]
\centerline{\includegraphics[width=15cm]{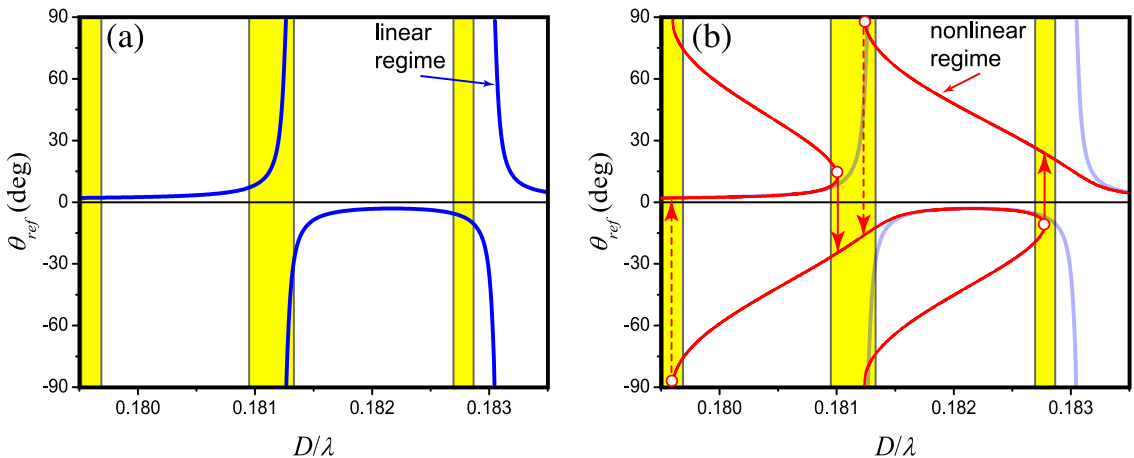}} \caption{(Color
online) Frequency dependences of the polarization azimuth $\theta_\text{ref}$
of the reflected field in the (a) linear and (b) nonlinear regime.
In the latter the input intensity $I_0$ is taken to be 10
kW/cm$^2$. Other parameters are as in Fig.~\ref{fig:fig2}. }
\label{fig:fig5}
\end{figure}

Expectedly, the angle of rotation of the polarization ellipse
($\theta$) in the nonlinear regime becomes a multi-valued function
too (Fig.~\ref{fig:fig5}). However, the ellipticity angle $\eta$ is
found to remain unchanged ($\eta=0$). The ambiguity of $\theta$ and
the associated bistability can be used to realize switching between
two different polarization states in the reflected light. The
frequency bands where this polarization instability takes place are
marked in Fig.~\ref{fig:fig5} as shaded areas. Switching between
different polarization states occurs as hopping between the two
stable branches when the frequency is varied. Such events are shown
in Fig.~\ref{fig:fig5} by solid and dashed arrows for increasing and
decreasing frequency, respectively.

\section{Time-domain simulation and polarization switching\label{sec:timedomain}}

The existence of bistability is a necessary but by no means a
sufficient condition for practical realization of bistable
switching, since the conditions needed to excite a particular branch
of the hysteresis loop remain to be determined
\cite{GrigorievNJP10,Grigoriev2010285}. Moreover, a hysteresis in
the frequency is of limited use because a continuous sweep of the
frequency would often be impractical in communication applications.
Therefore, in this section we demonstrate bistable polarization
switching in a magnetic nonlinear multilayer using time-domain
simulations.

To study signal evolution in the considered
structure, we start with a pair of wave equations for circularly-polarized
electric fields $E_{\pm}=E_x \pm i E_y$ in the form
\begin{eqnarray}
\frac{\partial^2 E_\pm}{\partial z^2} = \frac{\mu^\pm(z)
}{c^2}\frac{\partial^2 D_\pm}{\partial t^2}. \label{weq1}
\end{eqnarray}
Here $\mu^\pm=\mu_1^T\pm\alpha$ and $\mu^\pm=\mu_2$ in magnetic and
dielectric layers, respectively, $D_\pm=\varepsilon(z) E_\pm$ is the
electric displacement, the dielectric permittivity $\varepsilon (z,
I)=n (z, I)^2$ containing spatial periodic modulation and nonlinear
dependence on light intensity. The change of permittivity in the
each nonlinear layer $\Upsilon$ is described by Eq. (\ref{eq:eps}).

We solve Eqs. (\ref{weq1}) numerically using the finite-difference
time-domain (FDTD) method similar to that one developed in Ref.
\cite{Novitsky_PhysRevA.81.053814}. For convenience, we represent
field strengths as $E_+=A (t,z) \exp{[i( \omega t-kz)]}$ and $E_-=B
(t,z) \exp{[i( \omega t-kz)]}$, where $\omega$ is a carrier
frequency, $k=\omega /c$ is the wavenumber, and then solve equations
for the amplitudes $A (t,z)$ and $B (t,z)$. Finally, introducing
dimensionless arguments $\tau=\omega t$ and $\xi=kz$, the scheme of
calculation of the amplitude values at the mesh points
$(l\Delta\tau, j\Delta\xi)$ is as follows,
\begin{eqnarray}
A_j^{l+1}=[-\mu^+ a_1 A_j^{l-1}+b_1 A_{j+1}^l+b_2
A_{j-1}^l+(\mu^+ f-g) A_j^l]/\mu^+ a_2, \label{scheme1} \\
B_j^{l+1}=[-\mu^- a_1 B_j^{l-1}+b_1 B_{j+1}^l+b_2 B_{j-1}^l+(\mu^-
f-g) B_j^l]/\mu^- a_2, \label{scheme2}
\end{eqnarray}
where the auxiliary values are $ a_1=(n_j^{l-1})^2 (1-i\Delta\tau)$,
$ a_2=(n_j^{l+1})^2 (1+i\Delta\tau)$,
$b_1=\left(\Delta\tau/\Delta\xi\right)^2 (1-i\Delta\xi)$,
$b_2=\left(\Delta\tau/\Delta\xi\right)^2 (1+i\Delta\xi)$,
$f=(n_j^l)^2(2+\Delta\tau^2)$,
$g=2\left(\Delta\tau/\Delta\xi\right)^2+\Delta\tau^2$.

The stability of the algorithm is provided by the standard Courant
condition between the step intervals of the mesh; in our notation it
can be written as $\Delta\tau/\Delta\xi \leq n$. At the edges of the
calculation region, we apply the so-called absorbing boundary
conditions using the total field / scattered field (TF/SF) and the
perfectly matched layer (PML) methods \cite{taflove, anantha}.

\begin{figure}[htb]
\centerline{\includegraphics[width=15.0cm]{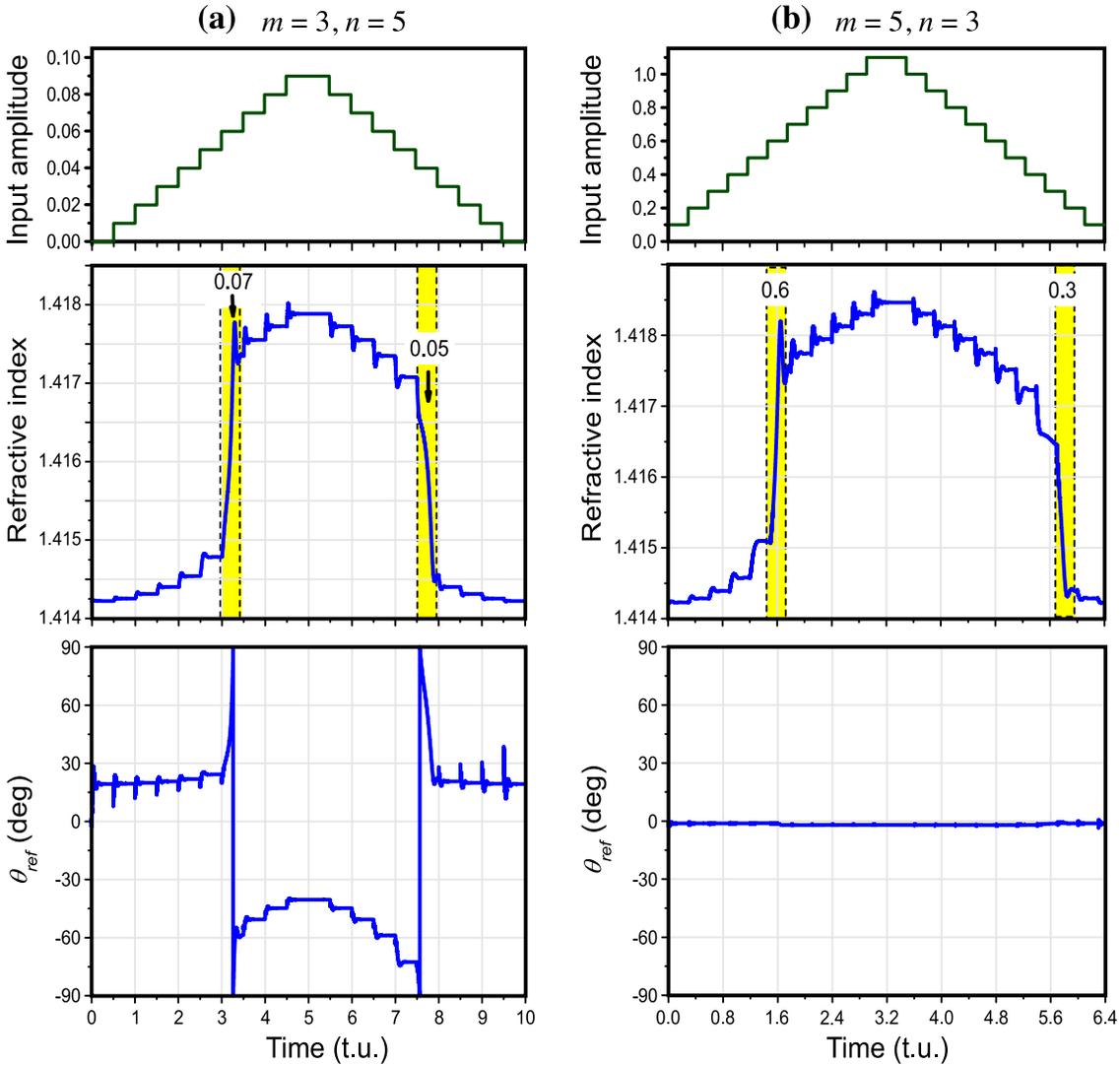}}
\caption{(Color online) The dynamics of the refractive index in the
middle of the cavity layer and the polarization azimuth of reflected
light for two configurations of asymmetric magnetophotonic crystal
for (a) $m=3$, $n=5$ and (b) $m=5$, $n=3$. The time is given in the
time units (t.u.) equal to $15000\lambda/c$, corresponding to 77.55
ps for the telecommunication wavelength of 1.55 $\mu$m. The
structure's parameters are as in Fig.~\ref{fig:fig2}. }
\label{fig:fig6}
\end{figure}

As previously, we take the asymmetric structure consisting of $8$
elementary units of two configurations of Section
\ref{sec:spectral}: with stronger ($m=3$, $n=5$) and weaker ($m=5$,
$n=3$) radiation localization inside the defect.  The permittivity
of material in the nonlinear layers is described by
Eq.~\eqref{eq:eps} with the linear part $\bar \varepsilon_2=2$ and
where $E_{in}(z)=E_{in}(t,z)$ is the actual value of the field.
Since we describe the fields using dimensionless amplitudes $A
(t,z)$ and $B (t,z)$, the normalized nonlinearity coefficient can be
fixed to be $\Bar {\Bar n}_2=\Bar {\Bar \varepsilon}_2/2 \sqrt{\bar
\varepsilon_2}=0.001$. These parameters are close to the case
considered in Section \ref{sec:spectral} for frequency-domain
calculations.
The incident radiation is assumed to be linearly polarized, so
that the amplitudes on the boundary of the system are equal,
$A(t,0)=B(t,0)$. The normalized thickness of the layers is $D/\lambda=0.183$.

To study the dynamics of polarization characteristics and achieve
switching, we performed the calculations with incident light
amplitudes gradually changing in a stepwise manner
(Fig.~\ref{fig:fig6}). The time unit (t.u.) contains $15000$ optical
cycles, i.e. $1$ t.u. $=15000 \lambda/c$. For the first
configuration ($m=3$, $n=5$, see Fig. \ref{fig:fig6}(a)), the
incident amplitudes $A$ and $B$ were changed in the range
$0.01$--$0.1$ (first increasing and then decreasing) with the step
$0.01$; every amplitude was kept constant during $0.5$ t.u. For the
second configuration ($m=5$, $n=3$, Fig. \ref{fig:fig6}(b)), the
amplitudes were changed in the range $0.1$--$1.1$ (the step is
$0.1$) after $0.3$ t.u.

Figures \ref{fig:fig6}(a) present clear evidence for the optical
bistability in the structure with stronger light localization
($m=3$, $n=5$). The refractive index in the middle of the double
layer (defect) and the polarization azimuth of reflected light
undergo abrupt switching two times at different input radiation
intensities. The corresponding amplitude values are $0.07$ (when
intensity is increasing) and $0.05$ (when  intensity is decreasing).
The polarization azimuth at these points also switches by almost
$90^\circ$, exhibiting an abrupt changes of sign.

The scenario turns out to be quite different in the weaker
localization regime ($m=5$, $n=3$) as seen in Fig.
\ref{fig:fig6}(b). Though the switching of the refractive index
still does occur (albeit at amplitudes an order of magnitude higher
-- $0.6$ and $0.3$, respectively), the corresponding switching of
the polarization state is negligibe. One can conclude
that the state of polarization remains almost unchanged: the
polarization azimuth does not change its sign and stays near zero
value.

\begin{figure}[htb]
\centerline{\includegraphics[width=15.0cm]{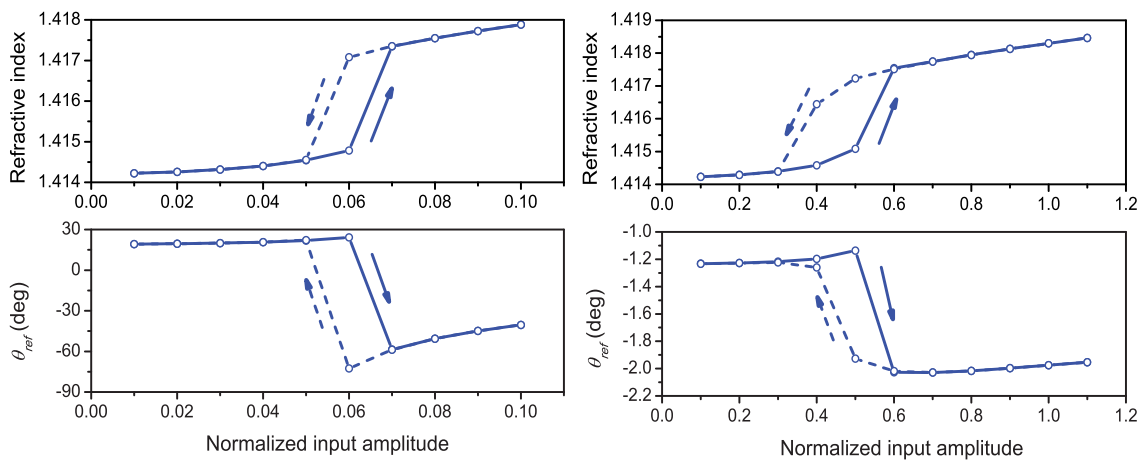}}
\caption{(Color online) The hysteresis loops for the refractive index in the middle of the cavity layer
($n_2^\text{cavity}$) and for the azimuth angle of reflected wave polarization ($\theta_\text{ref}$) for the two configurations in Fig.~\ref{fig:fig6}.
} \label{fig:fig7}
\end{figure}

The similarities and differences  between the two configurations are
highlighted in Fig.~\ref{fig:fig7}, which shows the explicit
hysteresis loops obtained from the calculations in
Fig.~\ref{fig:fig6} by plotting the relevant characteristics (the
refractive index inside the defect and the azimuth angle of the
output polarization) at the quasi-steady state after each input
intensity jump (after the transient oscillations have decayed)
versus the input intensity. It is seen that even though the loops
are quite narrow, they are undoubtedly present, confirming that
polarization bistability investigated in Section \ref{sec:spectral}
can actually be used for switching between almost orthogonal linear
polarizations.

\section{Conclusions\label{sec:conclusions}}

In summary, we have investigated spectral and polarization
properties of photonic multilayer structures combining both
magnetooptic properties (the Faraday effect), nonlinear optical
properties (the Kerr effect), and spatial inversion asymmetry. We
have considered a periodic structure with an asymmetrically placed
defect layer (a Fabry-Perot resonator with non-identical Bragg
mirrors) as the example geometry, and performed both
frequency-domain (nonlinear transfer-matrix) and time-domain (FDTD)
calculations.

Even in the linear regime, we have demonstrated that an interplay of
the Faraday effect and the spatial inversion asymmetry can give rise
to significant enhancement of optical non-reciprocity, resulting in
total $90^\circ$ polarization rotation of reflected light [see
Eq.~(\ref{eq:totalconversion})] for one direction of incidence only;
for the opposite direction of incidence, polarization conversion
effects were shown to vanish. This effect is explained by
considering a different coupling strength between the cavity mode
field and the incident field.

In the nonlinear regime, resonance bending and bistability in
nonlinear reflection spectra has been demonstrated. In a properly
designed multilayer (where output polarization is kept close to
linear throughout the stop band of the Bragg mirrors), this
bistability has been shown to result in bistability between distinct
linear polarization states of the reflected output light.

Finally, our time-domain numerical simulations have confirmed that
this polarization bistability can indeed be used for dynamic
polarization switching, clearly exhibiting hysteresis phenomena
(Fig.~\ref{fig:fig7}). Very strong dependence on the direction of
incidence of the input light has been demonstrated.

The results obtained can put forth new designs of both passive and
active multilayer-based optical devices, such as polarization
rotators, optical isolators, nonlinear polarization switches, and
nonlinear optical diodes. Additionally, it has been found that there
is a synergetic interplay between nonlinear, magnetooptical, and
spatial inversion, giving rise to polarization properties not
attainable otherwise. Investigating this interplay is a promising
subject of further studies.

\section{Acknowledgements}

This work was supported by the Ukrainian State Foundation for Basic
Research, project F54.1/004 and the Belarusian Foundation for Basic
Research, project F13K-009. One of us (S.V.Z.) wishes to acknowledge
partial financial support from the People Programme (Marie Curie
Actions) of the European UnionÕs 7th Framework Programme
FP7-PEOPLE-2011-IIF under REA grant agreement No. 302009 (Project
HyPHONE)

\bibliographystyle{tMOP}
\bibliography{Tuz_Novitsky}
\vspace{36pt}
\markboth{Taylor \& Francis and I.T. Consultant}{Journal of Modern Optics}

\label{lastpage}

\end{document}